# Period-doubling bifurcation in spatiotemporal mode-locked lasers

Xiaosheng Xiao,[1],*,† Yihang Ding,[2],† Shuzheng Fan,[1] Changxi Yang,[2],* Xiaoguang Zhang[1]

[1] State Key Laboratory of Information Photonics and Optical Communications, School of Electronic Engineering, Beijing University of Posts and Telecommunications, Beijing 100876, China.

[2] State Key Laboratory of Precision Measurement Technology and Instruments, Department of Precision Instruments, Tsinghua University, Beijing 100084, China.

**ABSTRACT** Period-doubling bifurcation is a universal dynamic of nonlinear systems, which has been extensively investigated in laser systems with a single transverse mode. This study presents an experimental observation and theoretical investigation of the period-doubling bifurcation in spatiotemporal mode-locked (STML) multimode fiber lasers. In the period-doubling state, it is observed that the pulse train modulation varies with the transverse-mode, and the output beam profile fluctuates rapidly and periodically. The numerical simulations conducted in this study are in good agreement with the experimental observations. Furthermore, a simple iterative model is proposed by considering the mode-dependent saturable absorption effect, and the experimental results can be qualitatively interpreted by this model. Based on these results, spatiotemporal saturable absorption is believed to be the key factor for the unique spatiotemporal characteristic of period-doubling bifurcation in multimode lasers. This study contributes to the understanding of the complex spatiotemporal dynamics in STML multimode lasers and to the discovery of novel dynamics in high-dimensional nonlinear systems.

Mode-locked lasers were typically operated in the form of a single transverse mode. Recently, spatiotemporal mode-locking was demonstrated in multimode fiber (MMF) lasers[1], which have subsequently attracted extensive attention. By using an MMF-based spectral filter, an all-fiber spatiotemporal mode-locked (STML) laser was realized[2]. Additionally, an STML fiber laser with spatiotemporal self-similar evolution was demonstrated[3]. Spatiotemporal mode-locking was achieved in a dispersion-managed MMF cavity[4], where pulse evolution was further achieved with Kerr-induced self-beam cleaning to realize single-mode output[5]. Recently, we demonstrated that spatiotemporal mode-locking could also be achieved in multimode cavities composed of MMFs with large modal dispersion, and a phenomenon of passive nonlinear auto-selection of single-mode mode-locking was observed[6]. Automatic spatiotemporal mode-locking was achieved in a multimode fiber laser by using genetic wavefront shaping[7].

A theoretical model, attractor dissection, was proposed to understand the mechanism of spatiotemporal mode-locking[8]. Several forms of spatiotemporal mode-locking that have no analogues to the (1+1)-dimensional system of single transverse-mode lasers have been identified. A model based on the (3+1)-dimensional cubic-quintic complex Ginzburg–Landau equation was proposed, using which soliton breathers and vortices were predicted in STML MMF lasers[9].

Mode-locked lasers provide an ideal platform to study the behavior of dissipative nonlinear wave dynamics[10]. Within the multimode cavities, there are much more complex nonlinear behaviors compared with single transverse-mode lasers. STML lasers are expected to support abundant nonlinear spatiotemporal dynamics because MMF cavities can provide a test bed with both temporal and spatial characteristics. Soliton molecules[11], multiple-soliton[12], wavelength-switching and hysteresis[13] have been observed in STML lasers. However, few STML states with spatiotemporal dynamics unique to multimode lasers were observed.

Bifurcation is a universal dynamic of nonlinear systems, which has been observed in physics, biology, chemistry, and even economics[14]. In laser systems, period-doubling bifurcation has been widely observed in semiconductor lasers[15], mode-locked solid-state lasers[16], mode-locked fiber lasers[17], etc. For the period-doubling state of mode-locked lasers, the output pulse train is not uniform and alters periodically. Period-doubling and pulsations in mode-locked lasers were theoretically investigated[18], and periodic pulsation was subsequently predicted in mode-locked fiber lasers[19]. Then period-doubling bifurcation was experimentally observed in a mode-locked fiber lase[20]. It was also observed that the soliton pulsation period can vary from a few to hundreds of roundtrips[21]. Numerous investigations have been conducted on period-doubling bifurcation and pulsation in mode-locked fiber lasers, e.g., period-doubling of multiple solitons[22], periodical pulsations in Tm-doped fiber lasers[23], and optical frequency combs based on a period-doubling mode-locked fiber laser[24]. A simple iterative model has been proposed to characterize the generation of period-doubling bifurcation[25,26].

In the abovementioned investigations, only one transverse-mode exists in the mode-locked lasers. However, in the STML MMF cavities, there are multiple transverse modes and a new spatial degree of freedom. Hence, unique spatiotemporal characteristics are expected in multimode mode-locked lasers. Multiple-period pulsation has been briefly mentioned in the supplementary materials of Ref. [1]. However, period-doubling bifurcation in STML MMF lasers has not yet been reported. In this study, we report on the experimental observation and theoretical investigation of period-doubling bifurcation in STML MMF lasers.

Our results indicate that in the period-doubling state, the nonlinear dynamics vary for different transverse modes. In the following, experimental observations are given first, then numerical simulations are introduced to validate the observed phenomena, finally a simple theoretical model is proposed to underline the physical insight of nonlinear dynamics in STML lasers.

We built two types of STML fiber lasers, the experimental setups of which are described in Section 1 of the supplementary material. The 1st laser is composed of passive MMF and active quasi-single-mode fiber, and the gain competition among the transverse modes can be neglected. The cavity structure is similar to that used in our previous study[11,12] and the first cavity mentioned in Ref. [1]. The core size (diameter of 10 μm) of the gain fiber is similar to those used in Refs. [2-5,13]. In this cavity, period-doubling bifurcation was experimentally observed and numerically validated, the details of which are explained in Sections 2 and 3 of the supplementary material. Although a variety of nonlinear dynamics (e.g., soliton molecule[11] and period-doubling bifurcation here) have been observed in cavities with quasi-single-mode gain fibers, the cavity modes are restricted compared with full-multimode cavities. Therefore, a 2nd fiber laser was built, composed of both multimode passive and active fibers, which is similar to that in our recent study[6]. About six and dozens of spatial modes can be supported in the active and passive MMFs, respectively. A nonlinear polarization rotation (NPR) technique-based saturable absorber (SA) is used for mode locking and to compensate the modal dispersion of the MMFs[6].

By adjusting the pump power and waveplates, spatiotemporal mode-locking with a uniform output can be easily achieved. However, period-doubling states can be achieved only in some sets of waveplates by further increasing the pump power. Typical uniform mode-locking is shown in Fig. 1(a), (c), and (e), which transitions to a period-doubling state by increasing the pump power, as shown in Fig. 1(b), (d), and (f). As shown in Fig. 1(a), the intensity of the output pulse train is uniform, and the pulse period is approximately 18 ns, which is in agreement with the fundamental repetition rate of 55.6 MHz. The corresponding radio frequency (RF) spectrum is illustrated in Fig. 1(c). There is no obvious sideband in this figure, implying that the output pulses are stable and uniform. The optical spectrum is shown in Fig. 1(e), and the multimode beam profile is displayed in the inset image. In contrast to the uniform mode-locking, the pulse train of the period-2 state, shown in Fig. 1(b), is not uniform; the pulse energy returns once every two roundtrips while maintaining the same interval. The RF spectrum is shown in Fig. 1(d), where there are two sidebands located at a frequency that is half of the cavity fundamental repetition frequency; this further validates the period-2 state. The optical spectrum and beam profile are shown in Fig. 1(f), which are slightly different from those in Fig. 1(e). When achieving the period-2 state, more period-doubling bifurcation is usually observed by further increasing the pump power. A typical state evolution is shown in Fig. 1(g) as pump power increases. In this figure, the STML state transitions from a uniform output to a period-2 state and then to a period-4 state.

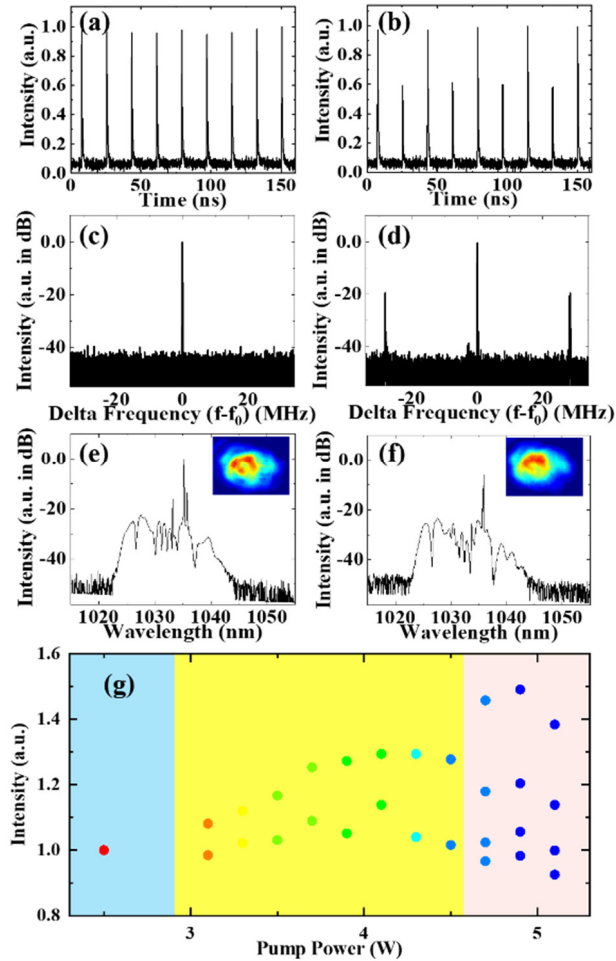

**Fig. 1** Experimentally observed typical STML states of **a,c,e** uniform output and **b,d,f** period-doubling (period-2 state). **a,b** Pulse trains recorded by oscilloscope, **c,d** the corresponding RF spectra, and **e,f** the corresponding optical spectra (inset: beam profile). **g** A typical evolution of period-doubling bifurcation with an increase in pump power. In **g**, the y-axis represents the values of the single-pulse energies in the output pulse train, measured by a photodetector and oscilloscope. For uniform mode-locking, there is one marker; for pulsation states, there are several markers.

To further investigate the spatiotemporal dynamics of the period-2 state, the output beam is spatially sampled and measured (setup shown in Fig. S1, similar to Ref. [12]). The extent of pulse fluctuation of the sampled pulse trains varies at different sampling positions; thus, the dynamics vary for different transverse modes. As an example, output pulse trains sampled at three different positions for the period-2 state of Fig. 1(b) are shown in Fig. 2(a). All sampled pulse trains still fluctuate with the period-2 state; however, they fluctuate differently, even out of phase. Figure 2(b) shows the extent of nonuniformity throughout the beam profile, and the corresponding beam profile is shown in Fig. 2(e), which is reconstructed as an intensity contour using 5 × 5 spatial samplings. As shown in Fig. 2(b), the intensity fluctuations of the sampled pulse trains vary extensively between 34%-188.5%. The experimental results show that the degree of fluctuation of the pulse train is different at different spatial positions (i.e., different transverse-mode components), which indicates that different transverse modes have their own dynamics. In addition, the single-shot beam profiles, which cannot be detected by the slow-response beam profiler, are reconstructed by the 5 × 5 spatial samplings, as shown in Fig. 2(c,d). A comparison between Figs. 2(c) and (d) validates that the output beam profile fluctuates quickly and periodically for the pulsation states in STML lasers.

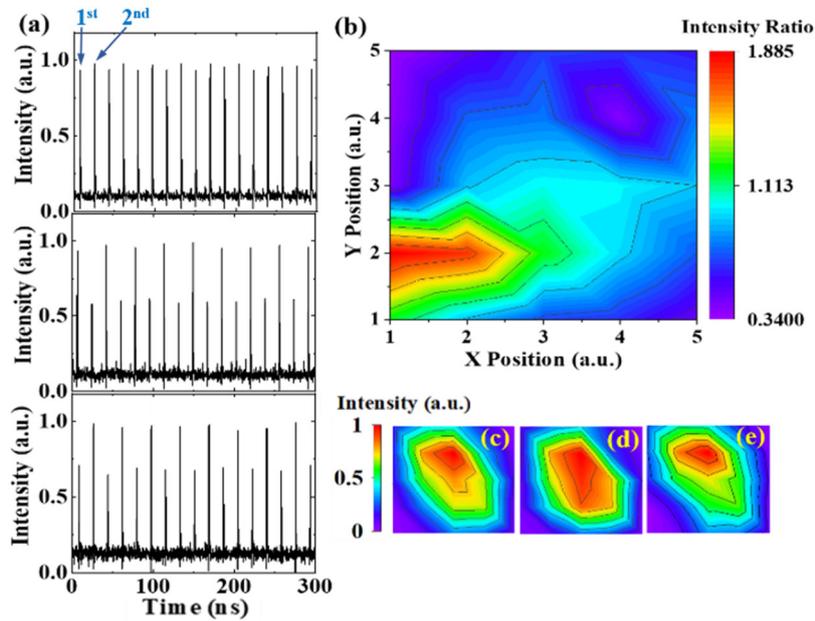

**Fig. 2 a** Spatially-sampled pulse trains at different positions; **b** contour of the intensity ratio for each sampled pulse train versus the sampling positions. The ratio is calculated by the intensity of the 1$^{st}$ pulse divided by that of the 2$^{nd}$ pulse labelled in **a**. **c-e** Contours of the intensity of sampled pulse trains: **c** intensity of the 1$^{st}$ pulse, **d** intensity of the 2$^{nd}$ pulse, and **e** average intensity of the 1$^{st}$ and 2$^{nd}$ pulses.

Other pulsation states with different periods are also observed in the cavity. Figure 3(a) shows the pulsation with odd periods (period-3) observed in the cavity. Further details of this state are presented in Fig. S8. Long period pulsations (e.g., with a period of 18 roundtrips) are also observed, as shown in Fig. 3(b).

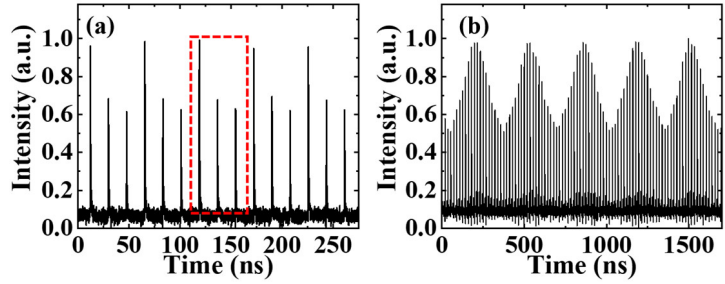

**Fig. 3** Two typical pulsation states with different periods. **a** Period-3 state; **b** long period pulsation.

Phenomena similar to those described above were also observed in the first cavity, as shown in Section 2 of the supplementary material. The experimental results suggest that the transverse-mode dependent period-doubling bifurcation is an intrinsic feature of MMF lasers. To confirm our observations, numerical simulations were conducted for both cavities. The generalized multimode nonlinear Schrödinger equation is used to simulate nonlinear pulse propagation in passive MMFs[27]. For the 1st cavity, the gain fiber is treated as a single-mode fiber, and a mode-resolved SA is considered. The experimental observations in the 1st cavity were confirmed by the simulations, the details of which are presented in Section 3 of the supplementary material. For the 2nd cavity, to model the gain competition among the transverse modes, the spatially dependent gain saturation is considered in the active MMF. An instantaneous spatiotemporal SA model is used, the transmission function of which is applied to the full spatiotemporal field composed of all modes. Details of the simulations are provided in Section 5 of the supplementary material.

By setting suitable parameters for the effective SA, etc., period-doubling bifurcation, i.e., the transition from stable uniform spatiotemporal mode-locking states to period-2 and -4 states, can be achieved by gradually increasing the saturation energy of the gain fiber. A typical simulated state evolution of the period-doubling bifurcation is shown in Fig. 4(a). For the period-2 state with a saturation energy of 10.6 nJ, the steady output pulse trains (represented by 4 pulses) are shown in Fig. 4(b,c). With the assistance of numerical simulations, the output pulses are mode-resolved, and we observe that the spatiotemporal pulses and the transverse mode components of the pulses alter periodically. Moreover, Fig. 4(c) shows that the pulse train fluctuations of different transverse modes are dissimilar, which is in good agreement with the experimental observations. An animation of the simulated period-doubling bifurcation, which occurs as the pump power increases, is shown in the supplementary movie. In addition, we also observed period-3 and multiple periods pulsation states using simulations, as shown in Fig. S9.

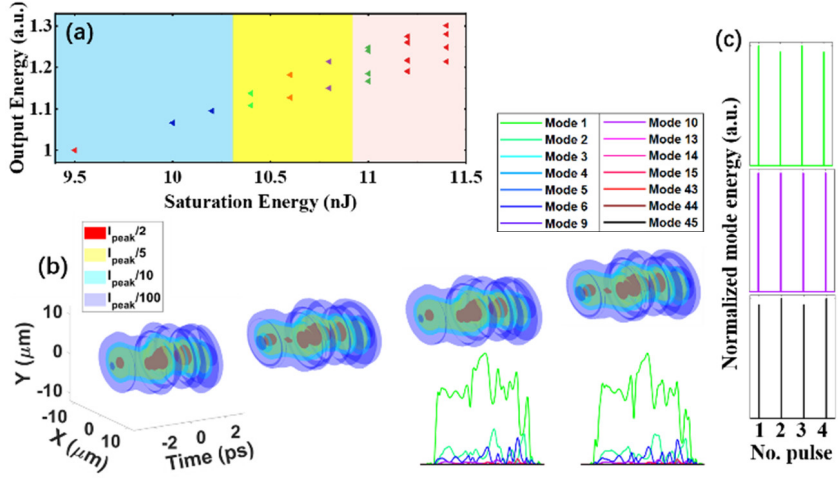

**Fig. 4** Simulation results of **a** period-doubling bifurcation for an increase in saturation energy in the STML MMF cavity, and **b,c** a steady output pulse train of the period-2 state with a saturation energy of 10.6 nJ. **b** Spatiotemporal intensity of four successive output pulses. Insets: the corresponding mode-resolved temporal shapes of the two differnt pulses; **c** mode-resolved output pulse trains of the 1st, 10th, and 45th modes.

We believe that the mode-dependent fluctuation in the pulsation state probably results from the spatial-dependent transmission (nonlinear loss) of the SA, especially considering the intensity-dependent NPR-based SA and the spatial distribution of the intensity in the multimode cavity. A very simple theoretical model is proposed to qualitatively simulate the nonlinear spatiotemporal dynamics of the MMF cavity. A sketch of the model is shown in Fig. 5(a), which is modified from the iterative model for single-mode fiber lasers[25,26,28]. Compared with the single-mode model, the modified model considers the transverse-mode-resolved SA and the mode coupling inside the cavity. Other mode-dependent effects, such as the gain competition among the transverse modes, are neglected. The multimode pulse is amplified as a whole; while the transmission of the SA is mode-dependent, as illustrated in Fig. 5(a). The pulse energy in successive transmitting through the gain medium and SA can be described by the following two equations, respectively:

$$E_{gt,n} = \frac{g_{net,0}}{1+E_{st,n}/E_{sat}} E_{st,n} \tag{1}$$

$$E^i_{s,n+1} = \frac{1}{2}\left[1-q'\cos\left(\pi E^i_{g,n}-\varphi_0\right)\right]E^i_{g,n}, \tag{2}$$

where $E_{st,n}$ and $E_{gt,n}$ are the total pulse energy (including all transverse modes) before and after the $n$-th passing through the gain medium, respectively;

$$E_{st,n} = \sum_{i=1}^{m} m_1^i E_{s,n}^i, \quad E_{g,n}^i = m_2^i E_{gt,n}$$

represent the mode coupling inside the multimode cavity, $E_{g,n}^i$ and $E_{s,n+1}^i$ represent the pulse energy distributed in the $i^{th}$ transverse mode before and after SA, respectively. For the coupling between the SA and the gain fiber, the pulse energy is re-distributed among the transverse modes, which is represented by the coupling coefficients of the $i^{th}$ mode, $m_1^i$ and $m_2^i$. For the gain medium, $E_{sat}$ represents the gain saturation energy, and $g_{net,0}$ denotes the small-signal net gain. For the NPR-based SA, a sinusoidal dependence of the transmittance on energy is considered[25]. $q'$ is a modified parameter for the modulation depth of the NPR, and $\varphi_0$ describes the linear bias of the NPR.

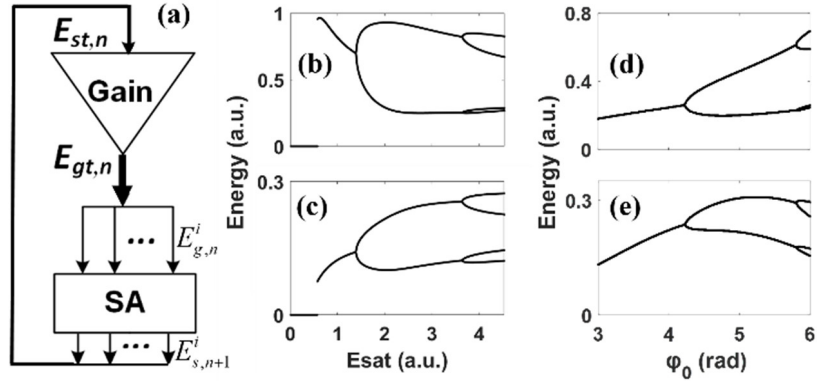

**Fig. 5 a** Theoretical iterative model and **b-e** examples calculated by the model. By increasing $E_{sat}$ (i.e., increasing the pump power), the calculated steady-state pulse energies distributed in the **b** 1$^{st}$ and **c** 3$^{rd}$ modes after the SA are presented. For a different phase bias of the NPR (varying $\varphi_0$), the pulse energies in the **d** 1$^{st}$ and **e** 3$^{rd}$ modes are presented. System parameters are $g_{net,0}$=6.5, $q'$=0.75, and $m_1^i$=$m_2^i$ ($i$=1~3) with $m_1^1$=4/7, $m_1^2$=2/7, $m_1^3$=1/7. For **b,c** $\varphi_0$=0, and for **d,e** $E_{sat}$=6. The initial condition is set to $E_{st,0}$=0.2, and the exact value of $E_{st,0}$ does not affect the steady state. SA: saturable absorber.

As a simple example of the model, we consider a cavity with three transverse modes. With an increase in pump power (by increasing $E_{sat}$), the calculated steady-state energies distributed in the 1$^{st}$ and 3$^{rd}$ modes are shown in Figs. 5(b) and (c), respectively. Figures 5(d) and (e) present the case of a varying phase bias of the NPR (by varying $\varphi_0$). Both cases show that the operation state experiences uniform output and subsequent period-doubling bifurcations with period-2 and -4. The evolution of Fig. 5(b,c) is similar to the experimental and numerical observations of the 2$^{nd}$ (full-multimode) cavity, shown in Figs. 1(g) and 4(a), respectively. The theoretical results of Fig. 5(d,e) are also in agreement with the observed period-doubling bifurcation in the 1$^{st}$ cavity by rotating the waveplate. Moreover, as shown in Figs. 5(b, c) or (d, e), by comparing the two modes, one can observe that the extent of the energy fluctuation varies with the transverse mode. This is in agreement with the observations in the two fiber cavities. The results of this simple model indicate that the mode-dependent SA is a key factor in the mode-dependent dynamics in the pulsation state of STML multimode lasers.

Saturable absorption (or nonlinear loss) is crucial for mode-locking, initiating and stabilizing the mode-locked states. In STML lasers, the SA exhibits unique effects, owing to its spatially-dependent saturable absorption. The SA was found to compensate the intracavity modal dispersion, thereby enabling multimode mode-locking in multimode lasers with large modal dispersion[6]. The SA was also found to be the key factor causing period-doubling bifurcation in conventional mode-locked lasers. In this study, period-doubling bifurcations with unique spatial-dependent dynamics were experimentally observed in both cavities with active quasi-single-mode fiber and active MMF. Although the gain fiber is considered as single-mode in the numerical simulations for the first cavity and in the theoretical model for the STML lasers, the experimental observations could be reproduced. These results indicate that the spatiotemporal SA is crucial for spatiotemporal dynamics in period-doubling states of STML lasers. Further novel nonlinear spatiotemporal dynamics may be identified by considering the spatiotemporal SA.

In our experiments of period-doubling bifurcation, if further tuning the cavity parameter (e.g., further increasing the pump power in the case of Fig. 1(g)), it usually transitions to a state for which the output pulse train is no longer periodic and the interval of the pulses remains the same. The output of this state is constantly in flux, and the dynamics require the use of more spatiotemporal measurement techniques and further investigations in the future.

The investigation of spatiotemporal dynamics in multimode lasers will benefit wider research fields. For example, space-division multiplexing has been extensively investigated in optical communications using MMFs[29], and Kerr solitons in different spatial modes have been generated in microresonators[30]. Our study will further enhance these multimode optical systems with circulating pulses. In the field of nonlinear science, many important nonlinear dynamical phenomena are essentially (3+1) dimensional. Thus, it is relevant to investigate (3+1) dimensional optical analogues in STML lasers, which are significantly more applicable to the important open questions in the field than conventional lasers.

In conclusion, period-doubling bifurcations have been reported in STML lasers with both active quasi-single-mode fibers and active MMF. The pulsation dynamics of different transverse modes were observed to be different. Numerical simulations for both cavities were conducted, which were in good agreement with the experimental observations. In addition, a simplified theoretical model was proposed to underlie the physical insight. The spatiotemporal SA is revealed to be a crucial factor for the spatiotemporal dynamics of period-doubling bifurcation in STML lasers. We believe that these results will enable the subsequent identification of more spatial-dependent and spatiotemporal dynamics in STML lasers, which have no analogues in lower dimensional nonlinear systems, including the conventional lasers operating in single transverse mode.

**Methods**

**Experiments.** The cavity configuration for both lasers is described in Suppl. Fig. S1. For the 1st cavity, the gain fiber is a 1.5-m-long double-cladding Yb-doped fiber (Liekki YB1200-10/125DC) with a core diameter of 10 μm and a numerical aperture (NA) of 0.08. The passive MMF is a 1.5-m-long, graded-index fiber (OM4 fiber, YOFC Corp.) with a core diameter of 50 μm and NA=0.2. For the 2nd cavity, the gain fiber is a 0.65-m-long Yb-doped MMF (Nufern LMA-YDF-20/125-9M) with a core diameter of 20μm and NA=0.08. The passive MMF is the same as that used for the 1st cavity and the length is 2.4 m. The free-space section includes waveplates, an isolator, a polarization beam splitter, and a spectral filter (bandpass filter with a center wavelength of 1030 nm and full width at half maximum (FWHM) of 10 nm). The gain fiber was pumped by a 976 nm laser diode. Self-starting STLM occurs with an appropriate cavity setup. More information about the setup and field measurements is provided in Section 1 of the Supplementary material and Refs. [1,6,11,12].

**Simulations.** For the 1st cavity, the quasi-single-mode gain fiber is approximated as a single-mode fiber[1,12]. The 6 lowest-order spatial modes of the passive MMF were considered. The effect of NPR is modeled by a mode-resolved lumped SA. The details of the simulation are presented in Section 4.1 of our previous paper[12] and Section 3 of the supplementary material. For both cavities, the generalized multimode nonlinear Schrödinger equation is used for the simulation of nonlinear pulse propagation in passive MMFs, which is solved by a massively parallel algorithm[27].

The simulation parameters of the 2nd cavity are based on those used in the experiments. The core diameter and NA of the active MMF are 20 μm and 0.08, respectively, and those of the passive MMF are 50 μm and 0.2, respectively, with a refractive index power of 2.08. The spatial modes, dispersions, and nonlinear coupling coefficients are calculated from the fiber specifications. All six spatial modes of the active MMF are considered. For the passive MMF that supports more than one hundred modes, different mode combinations are tested using numerous simulations, through which the modes that have taken effect are found. These modes are then considered in the subsequent simulations. To model the gain competition among the transverse modes, spatially dependent gain saturation is considered in the active MMFs. The gain spectrum exhibits a Gaussian profile with an FWHM of 40 nm. An instantaneous spatiotemporal SA model is used, the transmission function of which is applied to the full spatiotemporal field composed of all modes. A portion of the SA output is utilized as the laser output. Then, a spectral filter with a Gaussian profile and an FWHM of 10 nm is applied. We started the simulations at the input of the gain fiber with small pulses in all 6 modes. Simulations with other cavity parameters and simulation setups (e.g., different combinations of modes and initial inputs) were tested, and a similar conclusion can be achieved for a wide variety of parameters. More details on the simulations are presented in Section 5 of the Supplementary materials and Refs. [1,6,27].

**Acknowledgements**

This work was supported in part by National Key Scientific Instrument and Equipment Development Project of China under Grant 2014YQ510403, State Key Laboratory of IPOC (BUPT) (No. IPOC2020ZT02 and IPOC2019ZZ02), P. R. China, and National Natural Science Foundation of China (NSFC) (51527901, 61975090). The authors would like to thank Dr. Logan G. Wright for helpful discussion.

**Author contributions**

X.X. and Y.D. conceived the ideas. Y.D. and S.F. conducted the experiments. X.X. performed the simulations and proposed the theoretical model. X.X. helped performing the experiments. X.X., C.Y and X.Z. supervised the project. X.X., Y.D. and S.F. co-wrote the manuscript. All authors discussed the results and commented on the manuscript.

†These authors contributed equally to this work.

*Corresponding author. Email: xsxiao@bupt.edu.cn (X.X.); cxyang@mail.tsinghua.edu.cn (C.Y.)

Supplementary Information for

# Period-doubling bifurcation in spatiotemporal mode-locked lasers

Xiaosheng Xiao,[1]*† Yihang Ding,[2]† Shuzheng Fan,[1] Changxi Yang,[2]* Xiaoguang Zhang[1]

[1] State Key Laboratory of Information Photonics and Optical Communications, School of Electronic Engineering, Beijing University of Posts and Telecommunications, Beijing 100876, China.

[2] State Key Laboratory of Precision Measurement Technology and Instruments, Department of Precision Instruments, Tsinghua University, Beijing 100084, China.

The supplemental material including:

Section 1. Experimental setup

Section 2. Experimental results of the first cavity

Section 3. Numerical simulation and results of the first cavity

Section 4. Experimental observation of period-3 state in the second cavity

Section 5. Numerical simulation and additional results of the second cavity

Description of Additional Supplementary File (Movies S1)

## Section 1. Experimental setup

Two types of spatiotemporal mode-locked (STML) fiber lasers were built; the experimental setups of both lasers are illustrated in Fig. S1. A 980 nm multimode laser diode is pumped into a segment of the active fiber, which is spliced to a passive multimode fiber (MMF). The cavity also includes free-space optical devices for polarization control and optical coupling. An intracavity spectral filter (SF, bandpass filter with a 3-dB bandwidth of 10 nm) and a virtual spatial filter (induced by the optical coupling in the cavity) form a spatiotemporal filter [1,2]. The nonlinear polarization rotation (NPR) technique is used for mode locking and to enable the compensation of the large modal dispersion of the fiber [3].

The first cavity was composed of a quasi-single-mode gain fiber and passive graded-index MMF, and the cavity was similar to those in our previous work [2, 4], except that the length of the passive MMF is longer to induce more nonlinear interaction. The active fiber was a 1.5-m-long double-cladding gain fiber (Liekki YB1200-10/125DC with a 10 μm core and core and cladding numerical apertures (NAs) of 0.08 and >0.48, respectively). The passive MMF is a 1.5-m-long graded-index (GRIN) fiber (OM4 fiber, YOFC

Corp.) with a core diameter of 50 μm and NA=0.2. To excite many modes, the alignment of the passive GRIN fiber is offset from the gain fiber's center (by approximately 10 μm) during fiber splicing. Further details of the first laser are provided in Ref. [2,4]. The second cavity was composed of a 0.65-m-long step-index Yb-doped MMF (Nufern LMA-YDF-20/125-9M with a core diameter of 20 μm and NA=0.08) and a 2.4-m-long GRIN passive MMF (OM4, YOFC Corp.). About six and dozens of spatial modes can be supported in the active and passive fibers, respectively [3]. This cavity is similar to that utilized in our previous study [3], in which further details can be found.

To further investigate the spatiotemporal dynamics of the STML states, the output beam is spatially sampled and measured, similar to that in Ref. [4]. The sampling method is illustrated in Fig. S1; the output beams are directly coupled to two samplers simultaneously, one of which is fixed, and the other moves in the x-y plane. Then, the spatially sampled pulses are detected using two parallel 5 GHz photodetectors and recorded using an 8 GHz real-time oscilloscope. A total of 5×5 sampling points were measured. When comparing the pulse trains sampled at different positions, they can be synchronized by referring to the fixed sampled train.

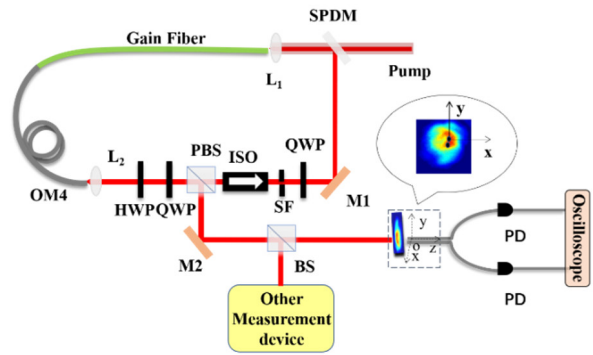

Fig. S1. Experimental setup of MMF laser. SPDM: short pass dichroic mirror; $L_1$ and $L_2$: collimating lens; Gain Fiber: quasi-single-mode or multimode ytterbium-doped fiber; OM4: passive GRIN multimode fiber; HWP: half-wave plate; QWP: quarter-wave plate; PBS: polarization beam splitter; ISO: isolator; SF: spectral filter; $M_1$ and $M_2$: mirrors; BS: beam splitter; PD: photodetector.

## Section 2. Experimental results of the first cavity

In this section, we describe the experimental observations in the 1st cavity, which is composed of active quasi-single-mode fiber and passive MMF. Using a pump power of 7.29 W and a suitable set of waveplates, stable spatiotemporal mode-locking with uniform output pulses can be easily achieved, as shown in Fig.

S2(a) and (b). Figure S2(a) shows a pulse train with a uniform intensity and an interval of ~16.5 ns, which is consistent with the 60.8 MHz fundamental repetition rate. Figure S2(b) shows the radio frequency (RF) spectrum. The signal-to-noise ratio at the fundamental repetition rate was measured at approximately 43 dB, confirming the uniform intensity pattern. By adjusting the waveplates, an operation state can be achieved whereby the pulse train is no longer uniform but the pulse intensity returns every two round trips while the pulses maintain the same interval, as shown in Fig. S2(c). For the measured RF spectrum, shown in Fig. S2(d), a new frequency component of 30.4 MHz, half of the fundamental cavity repetition, can be clearly observed. This further suggests that period-doubling bifurcation indeed occurs in the MMF laser. In the absence of a temperature-controlled laboratory environment, this period-2 state can last for several hours.

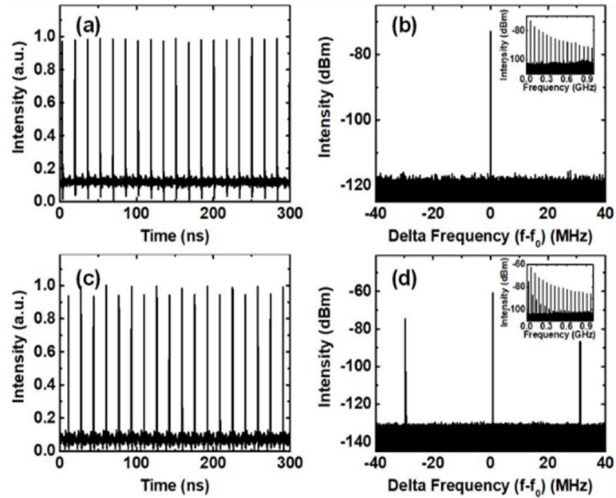

Fig. S2. Spatiotemporal mode-locking states with (a)(b) uniform output and (c)(d) a period-2 state. (a)(c) Pulse trains measured by the photodetector and oscilloscope; (b) (d) RF spectra in a 80 MHz span with a resolution of 510 Hz (inset: RF spectrum over a 1 GHz span).

To further confirm the period-2 state, dispersive Fourier transformation (DFT) measurements [5] were also performed, and the results (shown in Fig. S3) reveal that the shot-to-shot spectra also fluctuates periodically.

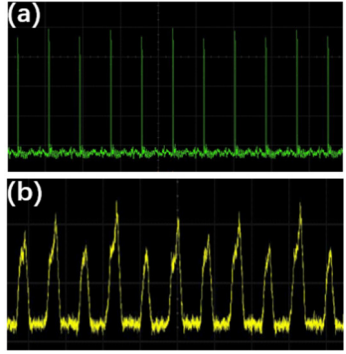

Fig. S3. Output pulse train of a period-2 state. (a) Directly measured using an oscilloscope; (b) the corresponding pulse train measured using the DFT technique, i.e., the output was transmitted through a spool of fiber before being measured by a fast photodetector and oscilloscope.

To further characterize the spatiotemporal mode-locking of the period-2 state, we utilize spatial sampling to analyze the output [4], as shown in Fig. S4. We observe several spatially sampled outputs by coupling only a part of the output beam into the samplers at different spatial positions. The sampled pulse trains at three different positions are presented in Fig. S4(a)–(c). Interestingly, the ratios of the two different intensity values, shown in Fig. S3(a)-(c), are 92.4%, 105%, and 72.9%, respectively. This indicates that the extent of the intensity fluctuations of the sampled pulses are markedly different at different spatial positions. The RF spectra of the three sampled pulse trains are shown in Fig. S4(e). According to this figure, the RF spectra are the same as that in Fig. S2(d) and superpose each other, which verifies the spatiotemporal mode-locking of the period-2 state. In addition, the entire output spectrum is displayed in Fig. S4(d) with the beam profile added as an inset image.

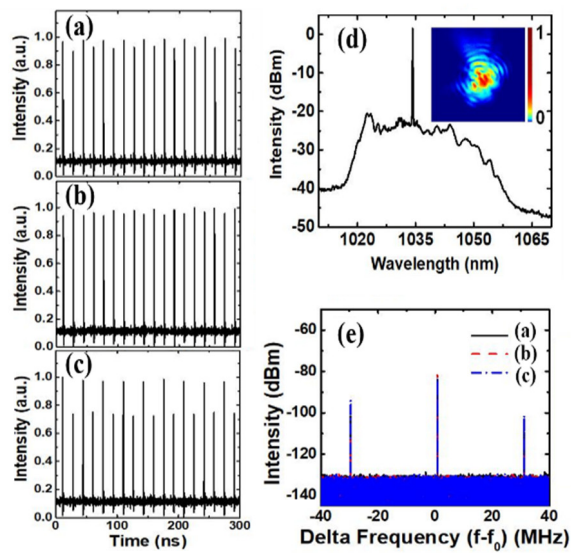

Fig. S4. Characterization of the STML period-2 state. (a)-(c) Different spatially-sampled pulse trains; (d) the optical spectra

of the entire output with the beam profile displayed in the inset image; (e) the corresponding RF spectra of the pulse trains of (a)-(c).

Additional to the period-2 mode-locking state, other states with different periods are also observed in the 1st cavity. The pulse trains of two typical states of multiple-period pulsations are shown in Fig. S5(a-b). A period-3 mode-locking state is obtained, as described in Fig. S5(a) and (c), by fixing the pump power at 7.29 W and rotating the intracavity half-wave plate. Long period pulsations (e.g., with a period of 17 round trips) can also be observed in this cavity, as shown in Fig. S4(b) and (d).

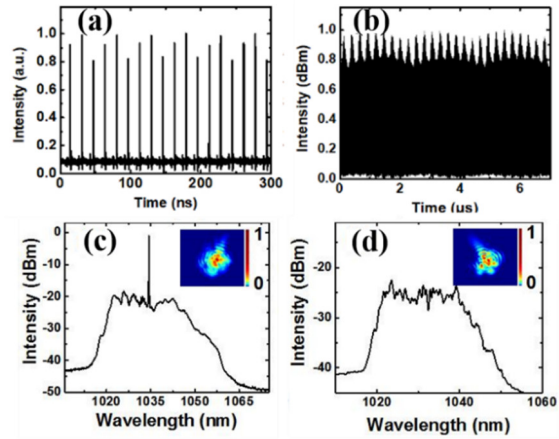

Fig. S5. Pulse trains of other STML states. (a) Period-3 state; (b) long period pulsations. (c) and (d) are the corresponding optical spectra of (a) and (b), respectively, with beam profiles displayed in the inset images.

Period-doubling bifurcation can be achieved by adjusting the rotation of the quarter-wave plate (QWP). As seen in Fig. S6, we rotated the QWP in one direction with a rotation angle increment of 2°, and recorded the output pulse energy and the operation state. The STML transitions from a uniform state to a period-2 and period-4 state (from region I–III).

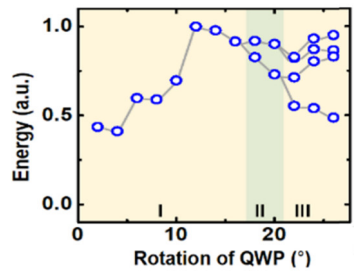

Fig. S6. Experimentally-measured period-doubling bifurcation in the 1st fiber laser, owing to gradual rotation of the QWP.

## Section 3. Numerical simulation and results of the first cavity

The numerical simulation of the 1st cavity is similar to that of the 1st cavity presented in Ref. [1], and the simulation, including the parameters used, is detailed in section 4.1 of our previous paper [4]. For the pulse propagation in the passive MMF, the generalized multimode nonlinear Schrödinger equation (GMMNLSE) is used [6], and the 6 lowest-order spatial modes of the passive MMF are considered. The quasi-single-mode gain fiber is approximated as a single-mode fiber [1]. The effect of nonlinear polarization rotation is modeled by a mode-resolved lumped saturable absorber (SA). The transmission function is $T = l_0[1 - q_0\cos(\pi|A_p|^2/P_{sat})]$ for each transverse mode $A_p$, where $l_0$ is the linear loss, $q_0$ is the modulation depth, and $P_{sat}$ is the saturation power of the SA; $A_p(z, t)$ is the electric field temporal envelope for the $p$-th transverse mode.

A typical simulation result is shown in Fig. S7, with the $q_0$ of the SA varying from 0.4 to 0.7. For each value of $q_0$, a steady mode-locking state was achieved. The energies of the steady output pulses are marked in Fig. S7(a). For uniform mode-locking, there is one marker; for pulsation states, there are several markers. As shown in Fig. S7(a), it is clear that period-doubling bifurcation occurs with an increase in $q_0$. As an example, the steady period-2 state with $q_0$=0.65 is shown in Fig. S7(b-e). For pulses with different energies, the transverse mode components are different, as shown in Fig. S7(c). Each transverse mode is also a period-2 state, as shown in Fig. S7(d,e). Comparing Fig. S7(d) and (e), one can observe that the extent of the fluctuation varies for different modes. These simulation results validate our experimental observations in the 1st cavity.

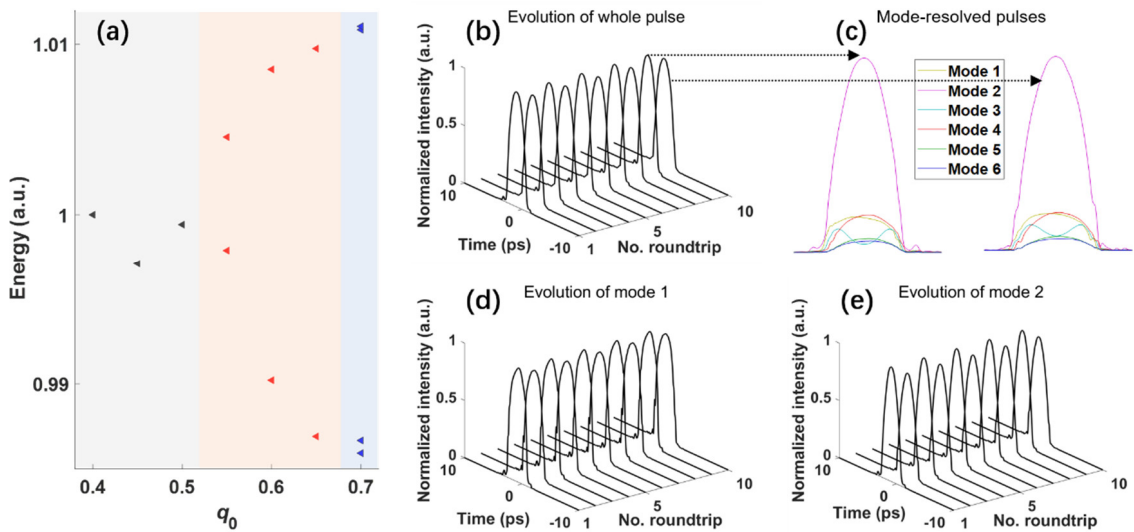

Fig. S7. Simulation results of (a) the period-doubling bifurcation with an increase of the modulation depth of the SA, $q_0$ in the 1$^{st}$ cavity, and (b-e) the steady output of the period-2 state with a $q_0$ of 0.65. (b) Output pulse train of ten successive output pulses, and (c) the corresponding mode-resolved temporal shapes of the two different pulses. (d) and (e) are mode-resolved output pulse trains with the 1$^{st}$ and 2$^{nd}$ modes given. In this simulation, for the SA, $l_0$=0.8 and $P_{sat}$=254 W; for the gain fiber, the small signal gain $g_0$=21 m$^{-1}$ and the saturation energy $E_{sat}$=1 nJ. Other parameters are detailed in Ref. [4].

## Section 4. Experimental observation of period-3 state in the second cavity

By appropriately adjusting the pump power and waveplates, the STML state with a uniform output is realized, and theperiod-2 state is then achieved by increasing the pump power. Occasionally, a stable period-3 state was found by further increasing the pump power and slightly turning the waveplate. One of the observed period-3 states is shown in Fig. 3(a). Herein, the RF and optical spectra of the period-3 state are given in Fig. S8 with the beam profile as the inset image.

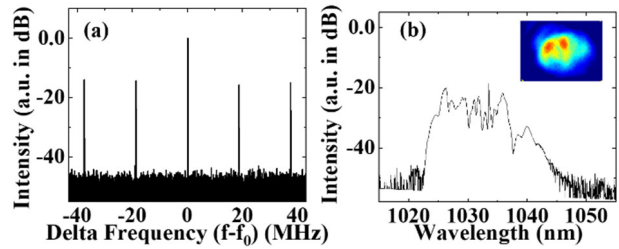

Fig. S8. The corresponding (a) RF and (b) optical spectra of the period-3 state of Fig. 3(a). Inset of (b): beam profile.

## Section 5. Numerical simulation and additional results of the second cavity

The simulation for the 1$^{st}$ cavity in Section 3 adopted a simplified model. Herein, for the 2$^{nd}$ cavity, more effects are considered in the simulations. The numerical simulation model and method used for the 2$^{nd}$ cavity are presented in the supplementary material of our previous paper [3]. To model the gain competition among the modes in the multimode gain fiber, the spatial dependent gain saturation is considered. The gain bandwidth was also considered. An instantaneous spatiotemporal SA model is used, the transmission function of which is applied to the full spatiotemporal field composed of all modes.

The GMMNLSE is used for the simulation of pulse propagation in passive MMFs [6]:

$$\partial_z A_p(z,t) = i(\beta_0^{(p)} - \Re[\beta_0^{(0)}])A_p - (\beta_1^{(p)} - \Re[\beta_1^{(0)}])\frac{\partial A_p}{\partial t} + i\sum_{n\geq 2}\frac{\beta_n^{(p)}}{n!}(i\frac{\partial}{\partial t})^n A_p +$$
$$i\frac{n_2\omega_0}{c}(1+\frac{i}{\omega_0}\partial t)\sum_{l,m,n}\left\{(1-f_R)S_{plmn}^K A_l A_m A_n^* + f_R A_l S_{plmn}^R \int_{-\infty}^{t} d\tau A_m(z,t-\tau)A_n^*(z,t-\tau)h_R(\tau)\right\}, \quad (1)$$

where $A_p(z,t)$ is the electric field temporal envelope for the $p$-th transverse mode, $\beta_n^{(p)}$ is the $n$th-order dispersion for the $p$-th mode, $S_{plmn}^K$ and $S_{plmn}^R$ are the nonlinear coupling coefficients for the Kerr and Raman effects, respectively, and $p$, $l$, $m$, and $n$ represent the numbers of the transverse modes. $f_R$ is the Raman contribution to the Kerr effect, $h_R$ is the Raman response of the fiber, $\Re$ denotes the real part only, and $n_2$ is the nonlinear index of refraction. In the simulations, the high-order (3$^{rd}$, 4$^{th}$, ...) dispersion, Raman ($f_R$ =0), and shock terms are neglected. For the active MMF, gain bandwidth and gain saturation are also considered with the GMMNLSE. To model the gain competition among transverse-modes, the model of the gain, considering the spatial dependence of the gain saturation, is [7]:

$$\frac{\partial A(x,y,z,\omega)}{\partial z} = \frac{1}{1+\int dt |A(x,y,z,\omega)|^2 / \tilde{I}_{sat}} \frac{g}{2} f(\omega) A(x,y,z,\omega),$$

where $A(x,y,z,\omega) = \sum_n F_n(x,y)A_n(\omega,z)$ with $F_n(x,y)$ be the mode profile of the $n$-th transverse-mode, $\tilde{I}_{sat}$ is the saturation time-integrated intensity, $g$ is the small signal gain, and $f(\omega)$ reflects the gain bandwidth. Additionally, $\tilde{I}_{sat} = E_s/A_{eff}$, where $E_s$ is the saturation energy of the gain fiber, and $A_{eff}$ is the effective mode area of the fundamental mode. A Gaussian profile with a full width at half maximum (FWHM) bandwidth of 40 nm was used to model the gain spectrum.

All 6 transverse modes are considered in the active MMF. For the passive MMF, approximately 120 modes can be guided. We are required to choose a part of the modes for the passive MMF, even though the massively parallel algorithm has been used to accelerate the simulation. Numerous simulations (e.g., different combinations of transverse modes in passive MMF and different initial inputs) were tested, and it was observed that stable spatiotemporal mode-locking can be achieved for a wide variety of parameters. Based on these simulations, fourteen modes that are relevant to the simulations were selected, i.e., the (1–6, 9, 10, 13–15, 43–45)-th modes. Simulations using more modes yielded similar results.

The spatiotemporal effect of the SA is applied to the full spatiotemporal field $A(x,y,z)$ [1]:

$$A_{out}(x,y,z) = A_{in}(x,y,z)\sqrt{1-\frac{q}{1+\left(|A(x,y,z)|^2/I_{sat}\right)}},$$

where $A(x,y,z)$ is the summation of all modes, $q$ is the modulation depth, $I_{sat}=P_{sat}/A'_{eff}$ is the saturation intensity of the SA, $P_{sat}$ is the saturation power, and $A'_{eff}$ is the effective mode area of the fundamental mode of the passive MMF.

The simulation parameters of the MMFs are consistent with those used in the experiments. For example, the spatial modes, dispersions, and nonlinear coupling coefficients are calculated based on the fiber specifications. For the gain fiber $g=33$ dB, and for the SA $q=1$. A lumped loss of 80%, including the loss owing to the output, is considered after the SA. Then, a spectral bandpass filter, centered at 1030 nm with a FWHM bandwidth of 10 nm, is applied to the field. For Fig. 4 and the supplementary movie, $P_{sat}$ of the SA is 16 kW. Other parameters of the cavity, e.g., the mode coupling, are the same as those in Fig. 1 of Ref. [3], if not specified.

Interestingly, in addition to the period-doubling bifurcation shown in Fig. 4, we also achieved period-3 pulsation states and multiple periods from the numerical simulations, as shown in Fig. S9. Figures S9(a) and S9(b) show the steady output pulse train and the mode-resolved temporal shapes of the period-3 state, respectively. Figure S9(c) shows the long-period pulsation state with a period of 12 roundtrips.

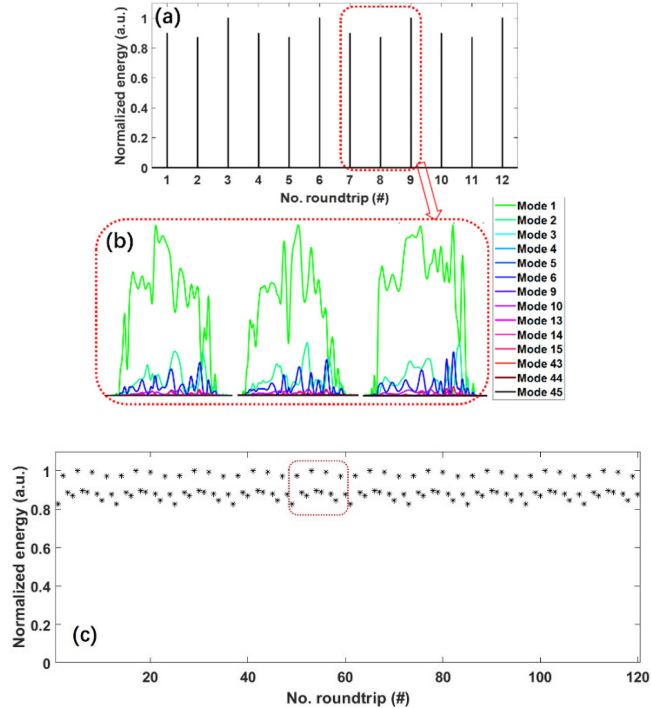

Fig. S9. Simulation results of (a,b) period-3 and (c) period-12 states in the 2$^{nd}$ cavity. (a) The pulse train and (b) the mode-resolved temporal shapes of the three different pulses in the priod-3 state. The saturation energy of the gain fiber is 9.25

nJ and the $P_{sat}$ of the SA is 14 kW. (c) Period-12 pulsation state. The saturation energy of the gain fiber is 9.15 nJ and the $P_{sat}$ of the SA is 11 kW. Other parameters are the same as those in Fig. 4.

## Description of Additional Supplementary File

**Movie S1. Intercavity evolutions of the output pulse as the saturation energy of the gain fiber $E_s$ increases from 10 to 11.2 nJ.** This movie is an animated version of the intercavity evolution of the output pulses, reflecting part of Fig. 4(a). First, $E_s$=10 nJ; the state starts from the initial input and approaches steady mode-locking with a uniform output. Then, the pump power increases (by setting $E_s$=10.6 nJ), and the state evolves to a period-2 state. Then, $E_s$ further increases to 11.2 nJ, and the state transitions to a period-4 state. The movie includes a mode-decomposed temporal pulse and spectrum, the beam profile, and the full spatiotemporal intensity of the pulse. The parameters are the same as those in Fig. 4.